# Predicting the Right Mechanism for Hypervalent Iodine Reagents by Applying Two Types of Hypervalent Twist Models: Apical Twist and Equatorial Twist

Tian-Yu Sun,[a,c] Kai Chen,[b,c]* Tingting You,[a]* Penggang Yin[a]

a. *Key Laboratory of Bio-Inspired Smart Interfacial Science and Technology of Ministry of Education, School of Chemistry, Beihang University, Beijing 100191, China. E-mail: youtt@buaa.edu.cn.*

b. *School of Chemistry and Chemical Engineering, Central South University, Changsha 410083, P. R. China. Email: kaichen@csu.edu.cn.*

c. *Lab of Computational Chemistry and Drug Design, State Key Laboratory of Chemical Oncogenomics, Peking University Shenzhen Graduate School, Shenzhen, 518055, China.*

## Abstract

Since the hypervalent twist followed by reductive elimination is a general reaction pattern for hypervalent iodine reagents, mechanistic studies about the hypervalent twist step provide significant guidance for experiments. Our previous work showed there are two types of hypervalent twist models, i.e. apical twist and equatorial twist. We applied both hypervalent twist models to explain the isomerization mechanism of two important electrophilic trifluoromethylating reagents, Togni I and Togni II. To the best of our knowledge, there are less detailed studies about the different twist modes between both reagents, which are important to predict the right reaction mechanism and especially, understand well the differences of reactivity and stability. Here, we successfully identified Togni II's isomerization pathway via equatorial twist, and suggested different hypervalent twist models should be considered to predict the right mechanisms of reactions with hypervalent iodine reagents.

Hypervalent iodine reagents are widely used in organic synthesis,[1] and the rearrangement of hypervalent bonds (also called intramolecular positional isomerization,[1] or Berry pseudorotation,[2] or hypervalent twist[3]) resulting in an exchange between the apical and the equatorial ligands in both $\lambda^3$- and $\lambda^5$-iodanes is important in explaining the mechanisms of hypervalent iodine reactions. Hypervalent twist followed by reductive elimination is the general reaction pattern for hypervalent iodine reagents, and experimental and theoretical studies had been carried out for the hypervalent twist step.[3-9]

In our recent work,[10] it was found that there are two types of hypervalent twist models, i.e. apical twist (from equatorial position to apical position) and equatorial twist (from one equatorial position to another equatorial position) for heterocyclic hypervalent iodine reagents (see **Scheme 1**). Apical twist and equatorial twist are also named as *in-plane* and *out-of-plane*, respectively, in Lüthi's work.[11] The differences of geometry structures and energy barriers between apical twist and equatorial twist are quite large. Distinguishing these two hypervalent twist models should benefit studies on the complicated mechanistic issues for hypervalent iodine reactions.

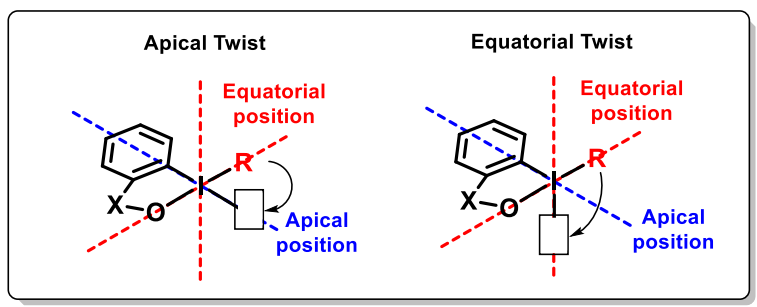

**Scheme 1.** Apical twist and equatorial twist for heterocyclic hypervalent iodine reagents.

In this work, density functional theory (DFT) calculations were conducted to study the reaction mechanisms of hypervalent iodine reagents. The M06-2X[12] method was chosen as the functional, and def2-TZVP[13] with *f*-functions was used as the basis sets. Structures were optimized in corresponding solvent with Truhlar's SMD method (Solvation Model based on the Quantum Mechanical Charge Density).[14] Our previous work showed that M06-2X/def2-TZVP is a reliable method for predicting the right mechanisms of reactions with hypervalent iodine reagents.[10]

Two heterocyclic hypervalent iodine reagents Togni I and Togni II with highly polarized $\lambda^3$ I-CF$_3$ bonds, first developed by Togni's group in 2006,[15-16] have been used widely as electrophilic trifluoromethylating reagents for the generation of new X-CF3 bonds (X = C, N, O, P, S, etc.).[17] According to our previous work,[18] since the *trans* influence of the I-O bond in Togni II is weaker than that in Togni I, the ability of Togni II to release $CF_3^+$ is stronger than that of Togni I. Moreover, Togni II could be activated by Lewis acids, Brønsted acids, or metals, through coordination with the carbonyl group.[17,19-20] Therefore, Togni II is used more frequently in organic synthesis.[17] Although Togni II is thermodynamically stable, just like Togni I,[6,21-23] the acyclic isomer of Togni II can be formed as a by-product during the trifluoromethylation reactions at high temperature (see **Scheme 2**).[6,23] Therefore, studies on the thermodynamics and kinetics of Togni II's isomerization might be helpful for avoiding the formation of acyclic by-product.

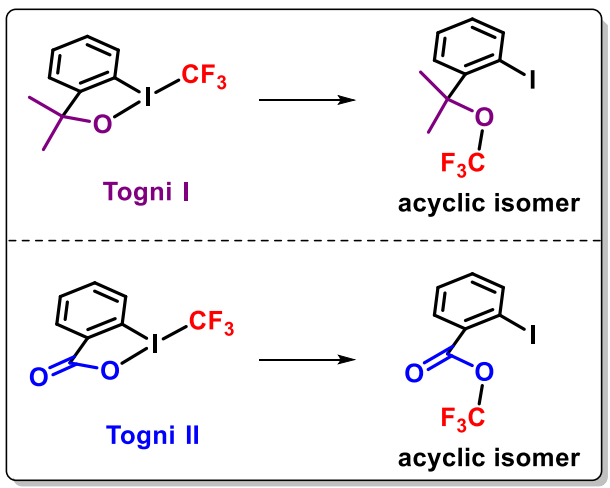

**Scheme 2**. The acyclic isomers of Togni I and Togni II as by-products.

Since the reaction performances of Togni I and Togni II are similar, we hypothesized the isomerization of Togni II underwent the same pathway as that of Togni I. As shown in **Figure 1**, Togni I goes through a two-step isomerization, i.e. the hypervalent twist (apical twist) and then the reductive elimination to form its acyclic isomer.[24] The energy barriers are too high to be accessed at room temperature. Interesting, our attempts to locate a similar apical twist transition state of Togni II failed, and to the best of our knowledge, there is still no reports about the apical twist transition state of Togni II.[11,25]

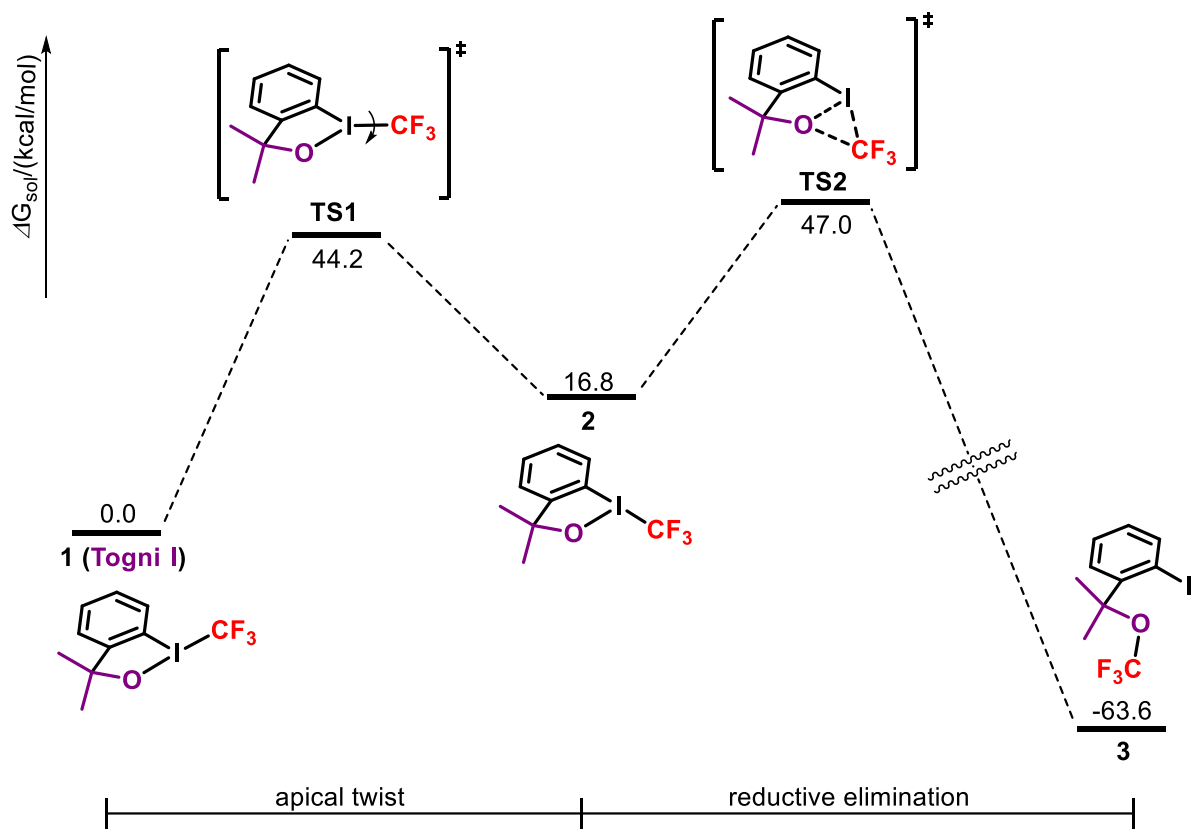

**Figure 1**. Togni I' isomerization PES via apical twist at M06-2X/def2-TZVP theoretical level. The free energy values are reported in kcal/mol.

Considering that the hetero-cleavage energy of I-CF$_3$ bond in Togni II is 40.9 kcal/mol, much lower than that in Togni I (64.5 kcal/mol) (see **Figure 2**), and meanwhile, the energy barrier of apical twist for Togni I is 44.2 kcal/mol, it is possible that the hetero-cleavage of Togni II's I-CF$_3$ bond occurred before the apical twist step finishes, if the energy barriers of apical twist are close between Togni I and Togni II. Then, the apical twist transition state of Togni II may not exist, as the CF$_3^+$ group will dissociate from iodine center when locating the apical twist transition state of Togni II.

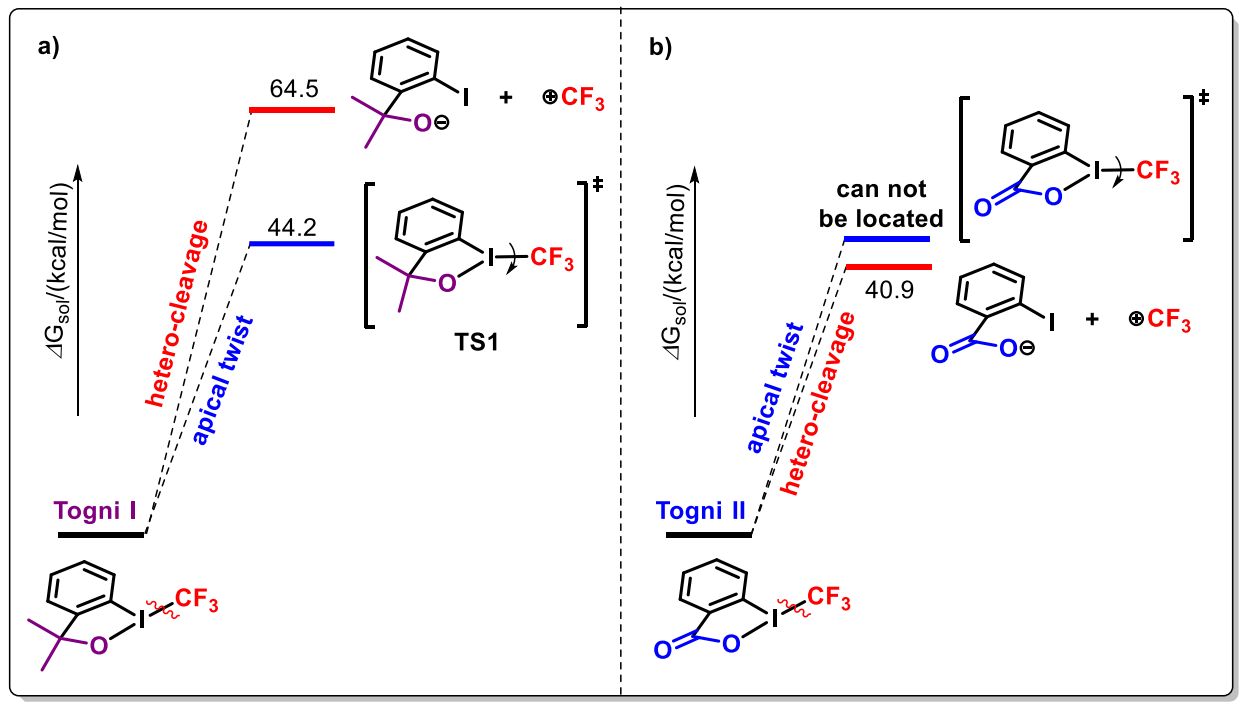

**Figure 2**. The hetero-cleavage energy of I-CF$_3$ bond in Togni I and Togni II. The free energy values are reported in kcal/mol.

Hence, it should be important to figure out the isomerization process of Togni II. According to our previous studies,[10] the hypervalent twist step is realized via an apical twist in Togni I but via an equatorial twist in IBX (an important hypervalent iodine oxidant).[26] As shown in **Figure 3a**, the CF$_3$ group of Togni I is predicted to move from the equatorial position towards the apical position. In **TS1**, the original I-CF$_3$ bond breaks along with the formation of new I-CF$_3$ bond by reacting with the anti-bond orbital of I-Ph bond. Since the I-CF$_3$ distance in **TS1** is lengthened by about 30%, and the *trans* influence of -Ph group is stronger than the linker -OCMe$_2$-,[27-28] the apical twist step **TS1** requires higher energy (> 40 kcal/mol). However, for the IBX-mediated oxidation of methanol, the equatorial twist step **TS3** is via the =O and -OCH$_3$ groups rotation around the I–Ph bond axis (see **Figure 3b**). The NBO analysis indicated the endocyclic I-OC(=O)R bond is highly ionic,[10,29] since the electron-withdrawing ability of -OC(=O)R is strong.[18] As a result, the I–OCH$_3$ and I=O distances almost unchanged in **TS3**, and the equatorial twist step requires lower energy (< 20 kcal/mol).

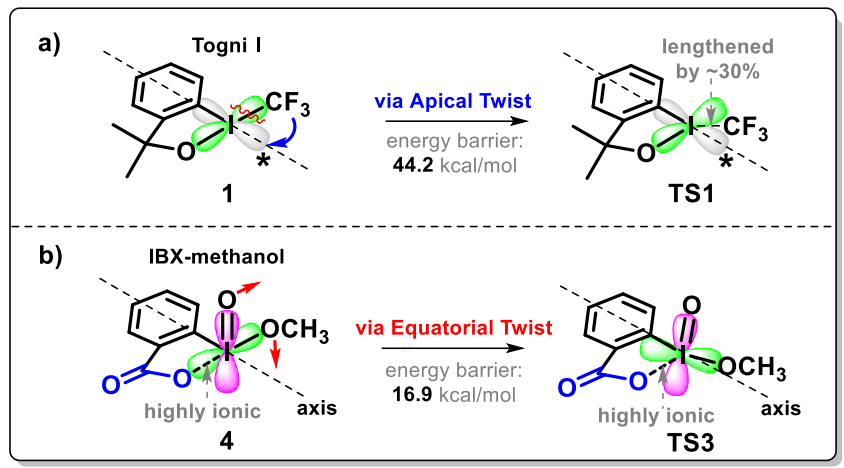

**Figure 3.** a) Schematic of apical twist for Togni I; b) Schematic of equatorial twist for IBX-methanol.

Inspired by the low energy barrier of equatorial twist, and the same linker -OC(=O)- between Togni II and IBX (see blue part in **Figure 2b** and **Figure 3b**), equatorial twist was considered for Togni II's isomerization (see **Figure 4**). To our delight, the equatorial twist step of Togni II (**TS4**) was successfully located and the energy barrier is just 16.4 kcal/mol. The two O atoms in linker -OC(=O)- are labelled as $O^1$ and $O^2$. In **TS4**, when the $CF_3$ group moves downward from one equatorial position to another equatorial position, the $O^1$ atom in I-$O^1$ bond moves in the opposite direction (upward) to keep the $O^1$-I-$CF_3$ three-center-four-electron (3c-4e) bond[1] until this bond broken. Meanwhile, the $O^2$ atom in the linker -OC(=O)- moves downward. After **TS4**, as shown in intermediate **6** in **Figure 4**, the $O^2$ atom is close to the $CF_3$ group. Then, via the reductive elimination step (**TS5**), which is the rate-determining step (RDS), the acyclic isomer product (**7**) could be obtained with the $O^2$-$CF_3$ bond formation. Gratefully, the isomerization pathway of Togni II was firstly disclosed, which is quite different from Togni I. The RDS's energy barrier of Togni II's isomerization is lower than that of Togni I's, and Koenen test showed that Togni II can be ignited by the flame of a match.[30]

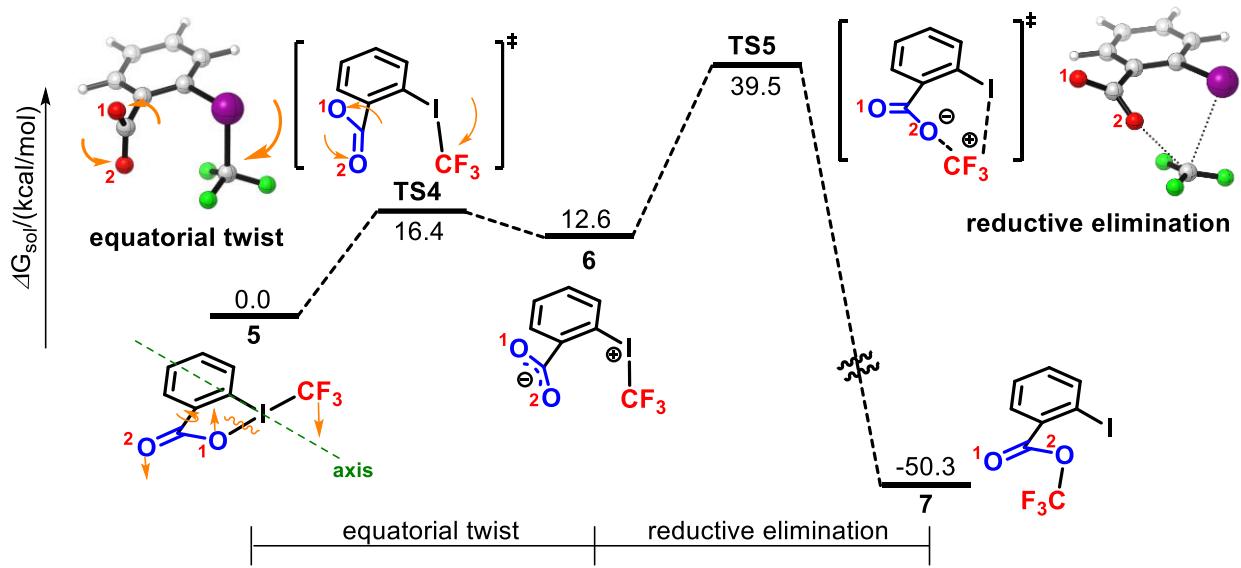

**Figure 4**. Togni II's isomerization PES via equatorial twist at M06-2X/def2-TZVP theoretical level. The free energy values are reported in kcal/mol.

Next, it would be interesting to revisit the equatorial twist pathway for Togni I. As shown in **Figure 5**, we have located the equatorial twist transition state **TS6** (45.9 kcal/mol) for Togni I, a little higher than the apical twist **TS1** (44.2 kcal/mol) of Togni I. Similar to Togni II, when the $CF_3$ group moves, the O atom in the I-O bond moves in the opposite direction (**TS6**). However, not like Togni II, there is only one O atom in the linker -$OCMe_2$- of Togni I (see purple part in **Figure 5**). For intermediate **8**, the O atom needs anticlockwise rotate further to get close to the $CF_3$ group (**TS7**). And if the O atom clockwise rotates to the previous I-O bond, the I-$CF_3$ bond will return to the original position (see **SI**). The steric repulsion of the two bulky methyl groups in the linker -$OCMe_2$- makes the equatorial twist step of Togni I (**TS6**, **8**, **TS7**) more difficult than that of Togni II (see **TS4** and **6** in **Figure 4**). Since the O atom in the linker -$OCMe_2$- of Togni I prefers to interact with I atom, the intermediate after **TS7** is the same as intermediate **2** in **Figure 1** with the formation of I-O bond. Intrinsic reaction coordinates (IRC) computations confirmed that **TS7** saddle-point structure indeed connects to intermediate **2** minima on the potential surface (see **SI**).[31-32] Finally, the reductive elimination step **TS2** of the equatorial twist pathway is also the same as that of the apical twist pathway for Togni I, and is also the RDS step. Due to the higher energy barrier and longer reaction path for the equatorial twist pathway of Togni I's isomerization, the apical twist is still the more feasible pathway. And this conclusion is consistent with Lüthi's PES scans.[11]

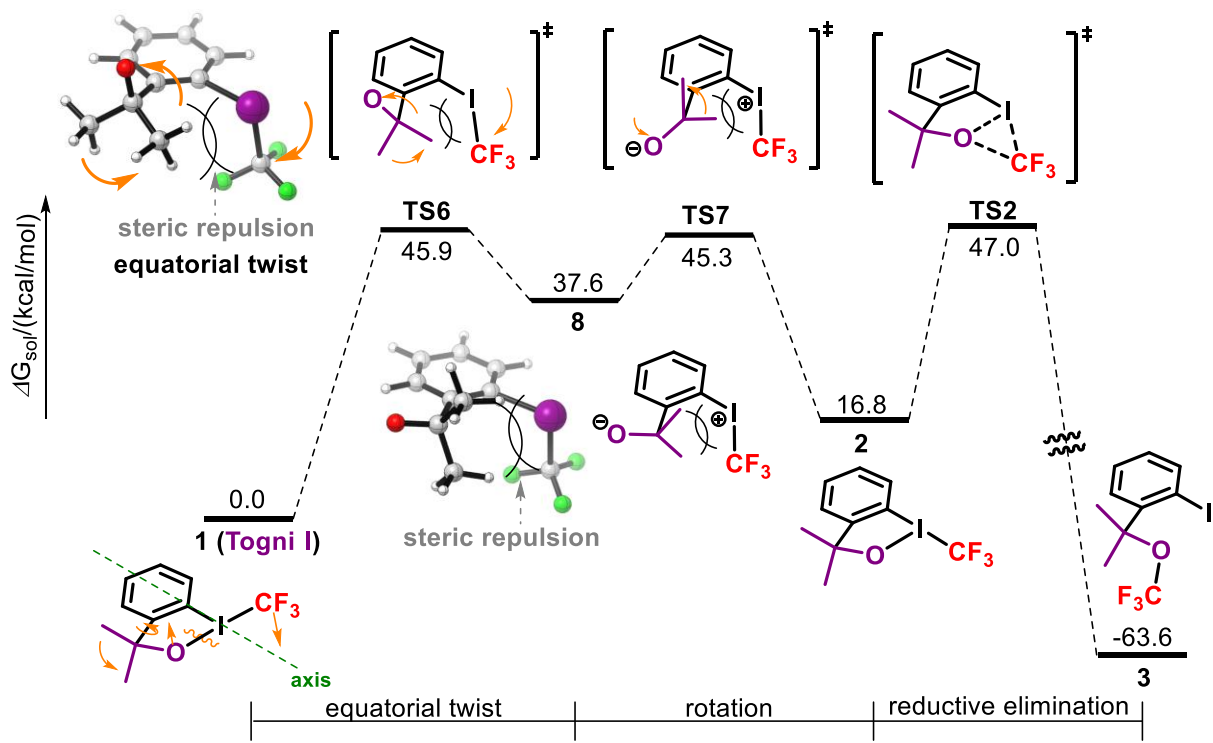

**Figure 5**. Togni I's isomerization PES via equatorial twist at M06-2X/def2-TZVP theoretical level. The free energy values are reported in kcal/mol.

As a summary, two types of hypervalent twist models, i.e. apical twist and equatorial twist, were adopted to study the mechanism of Togni II's isomerization. Since the heterolytic dissociation energy of I-CF$_3$ bond is low in Togni II, the I-CF$_3$ bond may undergo bond dissociation before the apical twist happens. Inspired by IBX's equatorial twist, we have successfully identified the equatorial twist pathway for Togni II. To our delight, the equatorial twist of Togni II is a lower-barrier step. Moreover, the second O atom in the linker of Togni II makes the following reductive elimination step feasible. However, owing to just one O atom and two bulky methyl groups in Togni I's linker, the equatorial twist pathway is more difficult than the apical twist pathway for Togni I. It is hopeful that our findings can facilitate the rational design of more active and stable hypervalent iodine reagents in the future, and these two hypervalent twist models (apical twist and equatorial twist) could help predict the right mechanisms of reactions with hypervalent iodine reagents.

## Conflicts of interest

There are no conflicts to declare.

## Acknowledgements

This work is supported by NSFC (31890774, 31890770), Guangdong NSF (2016A030310433), and Hunan NSF (2018JJ3868). Tian-Yu Sun acknowledges Li-Ying Pan for helpful discussions. KC acknowledges the National Supercomputer Center in Guangzhou for computer resources.

## Notes and references